\documentclass[aps,pra,preprint,groupedaddress,amsmath,amssymb,showpacs]{revtex4-1}
\usepackage{graphicx,amsmath,amssymb,epsfig}
\usepackage{dcolumn}
\usepackage{bm}
\usepackage{color}
\begin{document}

\title{\LARGE \sf Lossless Airy Surface Polaritons in a Metamaterial via Active Raman Gain}
\author{Qi Zhang$^{1}$, Chaohua Tan$^{1}$, and Guoxiang Huang$^{1,2, }$\footnote{gxhuang@phy.ecnu.edu.cn}}

\affiliation{$^{1}$State Key Laboratory of Precision Spectroscopy and Department of Physics, East China Normal University, Shanghai 200062, China\\
             $^{2}$NYU-ECNU Joint Physics Research Institute at NYU-Shanghai, Shanghai 200062, China
            }
\date{\today}
\maketitle

\noindent
\textbf{We propose a scheme to realize a lossless propagation of linear and nonlinear Airy surface polaritons (SPs) via active Raman gain (ARG). The system we suggest is a planar interface superposed by a negative index metamaterial (NIMM) and a dielectric, where three-level quantum emitters are doped. By using the ARG from the quantum emitters and the destructive interference effect between the electric and magnetic responses from the NIMM, we show that not only the Ohmic loss of the NIMM but also the light absorption of the quantum emitters can be completely eliminated. As a result, non-diffractive Airy SPs may propagate for very long distance without attenuation. We also show that the Kerr nonlinearity of the system can be largely enhanced due to the introduction of the quantum emitters
and hence lossless Airy surface polaritonic solitons with very low power can be generated in the system.}

In a seminar paper, Berry and Balaze~\cite{Berry} showed that a quantum-mechanical wavepacket with the form of
Airy function has the ability to resist dispersion and can freely accelerate without requiring any external
potential. It was argued later on that such wavepacket may be used to represent a nonrelativistic particle
falling in a  gravitational field, and hence the phenomenon discovered in Ref.~\cite{Berry} is related to
Einstein's equivalence principle~\cite{greenb}.

Since there is a similarity between the Schr\"{o}dinger equation in quantum mechanics and the Maxwell
equation in electrodynamics under a paraxial approximation, much efforts have been paid to the study of
Airy light beams in recent years due to their many attractive properties~\cite{Siv1,Siv2,Band}. Besides spatial-beam optics,
Airy beams have also been demonstrated for temporal optical pulses, spin waves, plasma, and electron
beams~\cite{Chong,Kim,Schn,Volo,Poly,Volo}.
In addition, some nonlinear effects of Airy beams have also been explored~\cite{Chen1,Fattal,Tsoy}.
Airy beams have a wide range of applications, including trapping, guiding, sorting of micro-objects,
manipulation of slow-light wavepackets in atomic gases, signal processing~\cite{Bau,Chen2,Zhang,Hang1,Rose}, and so on.

On the other hand, surface plasmon polaritons (SPPs), i.e. surface electromagnetic waves coupled to charge-density waves and propagating along the planar interface between a metal and a dielectric material, have attracted great attention~\cite{Maierbook,Saridbook}. SPPs have a field component decaying exponentially from metal-dielectric interface, thus can localize light within a subwavelength domain in the direction perpendicular the interface, making them ideal tools for enhancing light-matter interaction and hence for realizing many new types of nanoplasmonic devices~\cite{Gra,Stock,Ber,Kau,Hess}. However, the diffraction of SPPs in one of the directions in the metal-dielectric interface still exists.

Recently, Airy beams were introduced to a metal-dielectric interface as a technique for an effective control of SPPs~\cite{Sala}. The diffraction of SPPs in one of the directions in the interface, which is unavoidable in usual cases, can be eliminated by means of the non-diffractive property of Airy beams. Furthermore, some detrimental effects  resulted from the imperfection of the interface can be suppressed based on the self-healing characteristics of Airy beams. Such study~\cite{Sala} opened a new avenue for realizing nondiffracting SPPs in all transverse directions and stimulated many experimental efforts~\cite{Mino0,Zhu,Zhang2,Liu,Blec,Eps,Mino}. However, the Airy SPPs realized with such a scheme have a very short propagation distance due to the existence of large Ohmic loss inherent in metals, which severely limits their practical applications.

In this article, we propose a scheme for generating linear and nonlinear Airy surface polaritons (SPs) and realize their lossless propagation in an active metamaterial (for active optical metamaterials, see the recent review~\cite{Wue}). Different from previous studies~\cite{Sala,Mino0,Zhu,Zhang2,Liu,Blec,Eps,Mino,Wue}, the system we consider is a planar interface superposed by a NIMM and a dielectric where three-level quantum emitters are doped near the interface. By using the ARG from the quantum emitters and the destructive interference effect between the electric and magnetic responses in the NIMM, we show that not only the Ohmic loss of the NIMM but also the light absorption of the quantum emitters can be completely eliminated. As a result, non-diffractive Airy SPs obtained can propagate for a very long distance without attenuation and deformation. We also show that the Kerr nonlinearity of the system can be largely enhanced due to the introduction of the quantum emitters and hence lossless Airy surface polaritonic solitons propagating down the NIMM-dielectric interface with very low power can be realized.

\begin{figure}[htb]
\centering
\includegraphics[width=0.9\columnwidth]{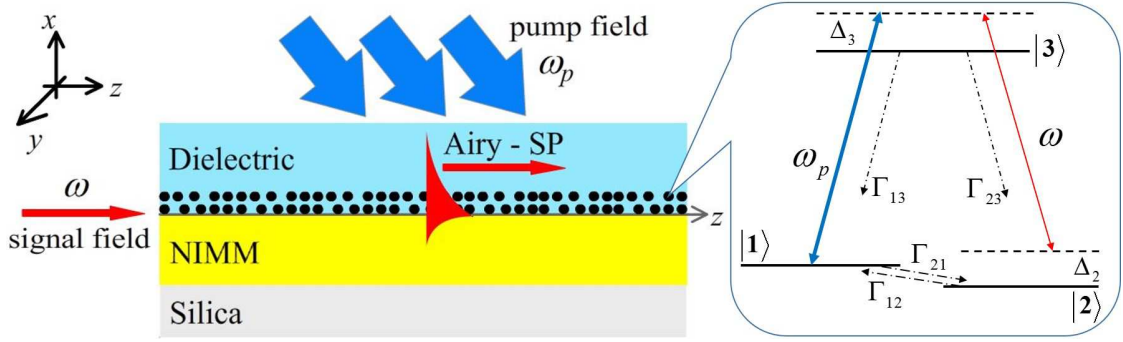}
\caption{\protect\footnotesize {\bf Model.}
Airy SP with angular frequency $\omega$ excited via ARG at the interface between a NIMM (in the region $x<0$) and a dielectric (in the region $x>0$). The lowest layer is a silica substrate. Inset: energy-level diagram and the ARG excitation scheme of the $\Lambda$-type quantum emitters (denoted by black dots) doped in the dielectric near the interface. $|j\rangle$ ($j=1,2,3$) are energy-levels of the quantum emitters and $\Delta_j$ $(j=2,3)$ are detuning. $\omega_p$  ($\omega$) is the angular frequency of the pump (signal) laser field, $\Gamma_{13}$  ($\Gamma_{23}$) is the rate of spontaneous emission from $|3\rangle$ to $|1\rangle$ ($|3\rangle$ to $|2\rangle$), $\Gamma_{12}$ ($\Gamma_{21}$) is the rate of incoherent population exchange from $|2\rangle$ to $|1\rangle$ ($|1\rangle$ to $|2\rangle$).
}\label{Fig1}
\end{figure}

\vspace{5mm}
\noindent\textbf{\Large \sf Results}\\
\noindent\textbf{\large \sf Model}.{\label{Sec2}}
We consider a system consisting of two superposed planar materials, i.e. a NIMM and a dielectric, with a planar NIMM-dielectric interface (Fig.~\ref{Fig1}).
The NIMM in the lower half-plane ($x < 0$) has frequency-dependent permittivity $\varepsilon_1$ and
permeability $\mu_1$, and the dielectric in the upper half-plane ($x > 0$) has
frequency-independent permittivity $\varepsilon_2$ and permeability $\mu_2$.
We assume that $\Lambda$-type three-level quantum emitters (e.g. atoms, quantum dots, rare-earth ions, denoted by black dots in the figure) are doped in the thin layer of
the dielectric near the interface, and interact with a pump field of angular frequency
$\omega_p$ and a signal field of angular frequency $\omega$; see the inset of Fig.~\ref{Fig1}.
$|j\rangle$ ($j=1,2,3$) represent the energy-levels of the quantum emitters.
$\Delta_2=(\omega_p-\omega)-(E_2-E_1)/\hbar$  ($\Delta_3=\omega_p-(E_3-E_1)/\hbar$) is two-photon
(one-photon) detuning, with $E_j$ the eigenenergy of the level $|j\rangle$. $\Gamma_{13}$ ($\Gamma_{23}$)
is the rate of spontaneous emission from $|3\rangle$ to $|1\rangle$  ($|3\rangle$ to $|2\rangle$),
$\Gamma_{12}$  ($\Gamma_{21}$) is the rate of incoherent population exchange from $|2\rangle$ to
$|1\rangle$ ($|1\rangle$ to $|2\rangle$). The three energy levels combined with the resonant pump
and signal fields constitute a typical ARG scheme discussed in Ref.~\cite{Huang1}.
SPs can be excited in the NIMM-dielectric interface~\cite{Mock} via an end-fire
coupling~\cite{Maierbook} for the signal field, with the pump field incident from
the above of the dielectric.

The system described above is similar to that employed in Refs.~\cite{Kamli,Moi,Kamli1,Sio},
where all-optical control of SPs through an excitation scheme of electromagnetically induced transparency (EIT) was
suggested. Differently, in stead of EIT, the excitation scheme of the quantum emitters employed
in our system is ARG. Contrary to the EIT scheme where signal field operates in an absorption mode, the central
idea of the ARG scheme is that the signal field operates in a stimulated Raman emission mode. It is just the
use of such emission mode that makes the Ohmic loss in the NIMM and the light absorption in the
quantum emitters eliminated and hence a robust propagation of the signal field realized, as shown below.

The SP propagation in the system is controlled by Maxwell equation describing electromagnetic
(EM) field and Bloch equation describing the quantum emitters. The Maxwell equation reads
\begin{equation}\label{Maxell00}
\nabla\times \left[ \nabla\times {\bf E} ({\bf r},t)\right]+\frac{1}{c^2}\,\frac{\partial^2 {\bf E}}{\partial t^2}=-\mu_0\frac{\partial^2}{\partial t^2} {\bf P}({\bf r},t)
-\mu_0 \frac{\partial}{\partial t} \left[\nabla\times {\bf M} ({\bf r},t)\right],
\end{equation}
where ${\bf E}({\bf r},t)$, ${\bf P}({\bf r},t)$, and ${\bf M}({\bf r},t)$ are intensity vectors of electric
field, electric polarization, and magnetization, respectively.
Throughout the text, for simplicity we assume all fields are continuous waves (CWs),
i.e. the dispersion effects
from both the host materials and the quantum emitters can be neglected. By
assuming the SP propagates in $z$-direction, we have
${\bf F}_{\alpha}({\bf r},t)={\bf F}_{\alpha}({\bf r})e^{i\left(\beta_{\rm a} z-\omega t\right)}+{\rm c.c.}$.
Here ${\bf F}_{\alpha}$ represents ${\bf E}_{\alpha}$, ${\bf P}_{\alpha}$, and ${\bf M}_{\alpha}$, and
${\bf F}_{\alpha}({\bf r})$ is a slowly-varying function of ${\bf r}$, with
$\alpha=1$ ($\alpha=2$) standing for the quantity in the NIMM region where $x<0$ (the dielectric region where $x>0$); $\beta_{\rm a}\equiv{\rm Re}(\beta_{\rm a}) + i{\rm Im} (\beta_{\rm a}$),  with ${\rm Re}(\beta_{\rm a})$ (real part) denoting the propagation constant and ${\rm Im}(\beta_{\rm a})$ (imaginary part) denoting the attenuation (if ${\rm Im}(\beta_{\rm a})>0$) or growth (if ${\rm Im}(\beta_{\rm a})<0$) of the SP during propagation. Because constitution relations are different in the NIMM  and the
dielectric regions, we reduce Eq.~(\ref{Maxell00}) in different regions for the convenience of later calculations.

In the NIMM region, we have
${\bf P}_1 ({\bf r},t)=\varepsilon_0 \left(\varepsilon_1-1\right){\bf E}_1({\bf r})e^{i\left(\beta_{\rm a} z-\omega t\right)}+{\rm c.c.}$
and $\nabla\times {\bf M}_1 ({\bf r},t)=-i\omega\varepsilon_{0}\varepsilon_1 \left(\mu_1-1\right){\bf E}_1({\bf r}) e^{i\left(\beta_{\rm a} z-\omega t\right)}+{\rm c.c.}$.
Then Eq.~(\ref{Maxell00}) reduces into
\begin{equation}
\nabla\times\left[\nabla\times\left({\bf E}_{1}({\bf r})e^{i\beta_{\rm a} z}\right)\right]-k_0^2\varepsilon_1 (\omega)\mu_1(\omega){\bf E}_{1}({\bf r})e^{i\beta_{\rm a} z}=0,
\label{ME for NIMM}
\end{equation}
with $k_0=\omega/c$. Note that $\varepsilon_1(\omega)$ and $\mu_1(\omega)$ are respectively the permittivity and permeability of the NIMM,
which can be parameterized by using the Drude model~\cite{Kamli,Moi,Kamli1} with
$\varepsilon_1 (\omega)=\varepsilon_{\infty}-\omega_e^2/(\omega^2+i\gamma_e\omega)$
and $\mu_1(\omega)=\mu_{\infty}-\omega_m^2/(\omega^2+i\gamma_m\omega)$,
where $\omega_e$ and $\omega_m$ are the electric and magnetic plasmon frequencies,
$\gamma_e$ and $\gamma_m$ describe the corresponding decay rates, and $\varepsilon_{\infty}$ and
$\mu_{\infty}$ are background constants, respectively.
Note that such permittivity and permeability can be obtained if the signal field is
normally incident into the NIMM designed by a periodical array of silver-based
double-fishnet structures of meta-atoms along $z$ direction (see Refs.~\cite{ZZhang,ZhangShuang,Shalaev1,Dolling,Shalaev} and related references cited in Ref.~\cite{Wue}).

In the dielectric region, one has ${\bf M}_2({\bf r},t)=0$ and
${\bf P}_2 ({\bf r},t)={\bf P}_2^{\rm dielectric} ({\bf r},t)+{\bf P}_{\rm emitter} ({\bf r},t)$, where
${\bf P}_2^{\rm dielectric} ({\bf r},t)=\varepsilon_0 \left(\varepsilon_2-1\right){\bf E}_2({\bf r}) e^{i\left(\beta_{\rm a} z-\omega t\right)}+{\rm c.c.}$  and
${\bf P}_{\rm emitter} ({\bf r},t)$ can be obtained by solving the Bloch equation of the quantum emitters (see below), given by
${\bf P}_{\rm emitter}({\bf r},t)=N_a\left[{\bf p}_{23}\sigma_{32}e^{-i\omega t}+{\rm c.c.}\right]$.
Here $N_a$ is the concentration of the emitters, ${\bf p}_{23}$  ($\sigma_{32}$) is electric-dipole matrix element (density matrix
element in interaction picture) related to the states $|2\rangle$ and $|3\rangle$.
Using these relations, Eq.~(\ref{Maxell00}) is reduced to the form
$\nabla\times\left[\nabla\times\left({\bf E}_{2}({\bf r}) e^{i\beta_{\rm a} z}\right)\right]+k_0^2\mu_2\left[\varepsilon_2{\bf E}_{2}({\bf r})e^{i\beta_{\rm a} z}+(N_a/\varepsilon_0){\bf p}_{23}\sigma_{32}\right]=0$.
To obtain $\sigma_{32}$, we must solve the Bloch equation~\cite{Boyd}
$i\hbar \left(\frac{\partial}{\partial t}+\Gamma\right)\sigma=\left[H_{\rm int},\sigma\right]$,
with $\sigma$ the $3\times 3$ density matrix (with matrix element $\sigma_{jl}$),
$H_{\rm int}$ the interaction Hamiltonian of the quantum emitters, and $\Gamma$
the $3\times 3$ relaxation matrix describing the spontaneous
emission and dephasing of the system. The explicit form of the Bloch equation and the result of
$\sigma_{32}$ obtained through solving the Bloch equation are presented in Methods. As a result, we have
\begin{equation}
\nabla\times\left[\nabla\times\left({\bf E}_{2}({\bf r}) e^{i\beta_{\rm a} z}\right)\right]-k_0^2\mu_2
\left[\left(\varepsilon_2+\chi_a^{(1)}\right)+
\chi_a^{(3)}|{\bf E}_{2}({\bf r})|^2e^{-2{\rm Im}(\beta_{\rm a})z}\right]{\bf E}_{2}({\bf r})e^{i\beta_{\rm a} z}=0,
\label{ME for dielectric1}
\end{equation}
where
\begin{equation}\label{CHI}
\chi_{a}^{(1)}=\frac{N_a |{\bf p}_{23}|^2}{\varepsilon_0 \hbar}a_{32}^{(1)},\,\,\,\,\,
\chi_{a}^{(3)}=\frac{N_a |{\bf p}_{23}|^4}{\varepsilon_0 \hbar^3}a_{32}^{(3)},
\end{equation}
are respectively the first-order and the third-order optical susceptibilities contributed
by the quantum emitters, where the definitions of $a^{(1)}_{32}$ and $a^{(3)}_{32}$ can be found in Methods.

Since the oscillating frequency of the pump field is different from that of the signal field,
the pump field has no contribution to the boundary conditions (BCs) of the signal-field
envelopes at the NIMM-dielectric interface. Thus the BCs read
${\bf e}_x\times ({\bf E}_2 ({\bf r})-{\bf E}_1 ({\bf r})\,)=0$
and ${\bf e}_x\cdot ({\bf D}_2 ({\bf r})-{\bf D}_1 ({\bf r})\,)=0$
(where ${\bf e}_x$ is the unit vector along $x$-direction), i.e.
\begin{eqnarray}\label{B.C.}
&& E_{1z}({\bf r})|_{x=0}=E_{2z}({\bf r})|_{x=0},\,\,\,\,E_{1y}({\bf r})|_{x=0}=E_{2y}({\bf r})|_{x=0},\\
&& \varepsilon_0\varepsilon_{1}E_{1x}({\bf r})|_{x=0}=\left[\varepsilon_0
\varepsilon_{2}E_{2x}({\bf r})e^{i\beta_a z}+N_a\left({\bf p}_{23}\sigma_{32}\right)_x\right]|_{x=0},\label{B.C.2}
\end{eqnarray}
where $F_{\alpha j}({\bf r})$ represents the $j$-component ($j=x,y,z$) of ${\bf F}_{\alpha}({\bf r})$ ($\alpha=1,2$).
Note that Eq.~(\ref{B.C.2}) is a nonlinear BC since $\sigma_{32}$ depends on ${\bf E}_2 ({\bf r})$ nonlinearly.

\vspace{5mm}
\noindent\textbf{\large \sf  SP solution and linear dispersion relation}.
Now we present the propagating modes of SPs in the system. Different from metal-dielectric interfaces, our
system allows both TE and TM modes. Here we concentrate on the TM mode, which has the form
${\bf E}_{\alpha}({\bf r})={\bf e}_xE_{\alpha x}(x)+{\bf e}_zE_{\alpha z}(x)$.
By solving Eqs.~(\ref{ME for NIMM}) and (\ref{ME for dielectric1}) under BCs (\ref{B.C.}) and (\ref{B.C.2}) in linear
level, we obtain
\begin{eqnarray}
&& {\bf E}_1({\bf r})=
\left({\bf e}_z-{\bf e}_x\frac{i\beta_{\rm a}}{k_1}\right)A e^{k_1x},\label{EM01}\\
&& {\bf E}_2({\bf r})=
\left({\bf e}_z+{\bf e}_x\frac{i\beta_{\rm a}}{k_2}\right)A e^{-k_2x},\label{EM02}
\end{eqnarray}
with $k_{1}=\left(\beta_{\rm a}^2-k_0^2\varepsilon_{1}\mu_{1}\right)^{1/2}$ and $k_{2}=\left[\beta_{\rm a}^2-k_0^2\mu_{2}\left(\varepsilon_{2}+\chi_{\rm a}^{(1)}\right)\right]^{1/2}$,
where $A$ is a constant. The linear dispersion relation (propagation constant) reads
\begin{equation}
\beta_{\rm a}(\omega)=\frac{\omega}{c}\sqrt{\frac{\varepsilon_{1}(\varepsilon_{2}+\chi_{\rm a}^{(1)})
[(\varepsilon_{2}+\chi_a^{(1)})\mu_1-\varepsilon_{1}\mu_2]}{(\varepsilon_{2}+\chi_{\rm a}^{(1)})^2
-\varepsilon_{1}^2}}.
\label{DR1}
\end{equation}

We first discuss the case where the emitters are absent (i.e.  $N_{\rm a}=0$ and hence $\chi_{\rm a}^{(1)}=0$). In this case, Eq.~(\ref{DR1}) reduces to
\begin{equation}
\beta_{\rm}(\omega)=\frac{\omega}{c}\sqrt{\frac{\varepsilon_{1}\varepsilon_{2}
(\varepsilon_{2}\mu_1-\varepsilon_{1}\mu_2)}{\varepsilon_{2}^2
-\varepsilon_{1}^2}}.
\label{DR0}
\end{equation}
For illustrating the character of the above result, we consider a realistic physical system with
a silver-based NIMM~\cite{Dolling}. The parameters for the permittivity are given by~\cite{Kamli,Dolling}
$\gamma_{e}=9\times10^{13}\,{\rm s}^{-1}$, $\omega_{e}=1.37\times10^{16}\,{\rm s}^{-1}$, $\varepsilon_{\infty}=1.7$, where $\gamma_{e}$ is assumed to be three larger than that of bulk silver~\cite{Shalaev}. The parameters for the permeability are given by $\gamma_{m}=5\times10^{9}\,{\rm s}^{-1}$, $\omega_{m}=3.15\times10^{15}\,{\rm s}^{-1}$,
and $\mu_{\infty}=1.7$, within the reasonable value scope~\cite{Kamli}.

\begin{figure}
\centering
\includegraphics[width=0.75\columnwidth]{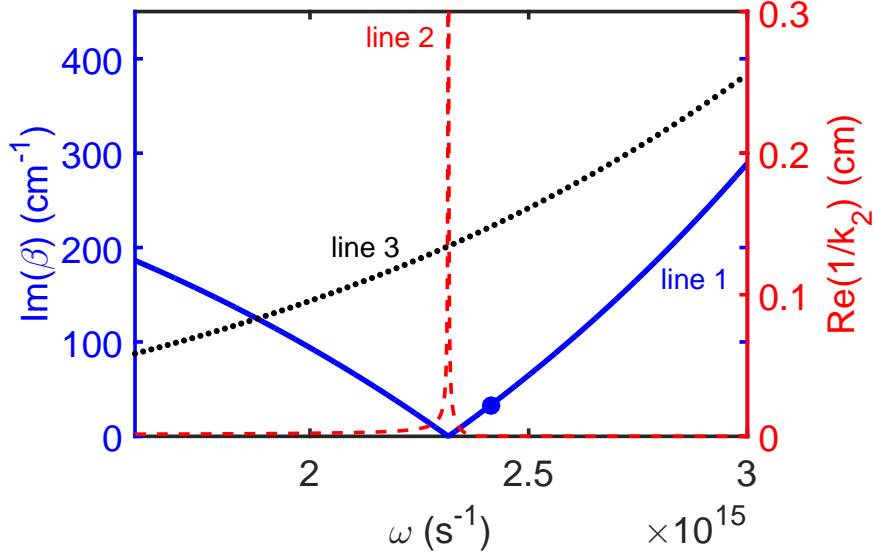}
\caption{\protect\footnotesize {\bf Linear dispersion relation of SP.}  ${\rm Im}(\beta)$ (line 1; blue solid line) and ${\rm Re}(1/k_{2})$ (line 2; red dashed line) of the SP in the NIMM-dielectric interface as a function of $\omega$.
The black dotted line (line 3) is the ${\rm Im}(\beta)$  of the SP in a metal-dielectric interface.
The large blue solid circle on line 1 corresponds to the selected signal-field frequency
$\omega=\omega_s=2\pi\times3.84\times10^{14}\,{{\rm s}^{-1}}$.}
\label{FIG2}
\end{figure}

Line 1 (blue solid line) and line 2 (red dashed line) in Fig.~\ref{FIG2}
show ${\rm Im}(\beta)$ and ${\rm Re}(1/k_{2})$ of the SP excited in the NIMM-dielectric interface as a function of $\omega$, respectively. When plotting the figure, we have chosen the dielectric with $\varepsilon_{2}=2.5$ and $\mu_{2}=1$. We see that ${\rm Im}(\beta)$ is nearly vanishing at $\omega=\omega_{\rm cri}
=2\pi\times3.69\times10^{14}\,{{\rm s}^{-1}}$, which means that the Ohmic loss at
$\omega_{\rm cri}$ is largely suppressed. The reason for the suppression of the Ohmic loss
is contributed by the destructive interference of the electric and magnetic responses
because $\varepsilon_1$ and $\mu_1$ in the NIMM can be simultaneously negative~\cite{Kamli}.
For comparison, ${\rm Im}(\beta)$ for a metal (silver)-dielectric interface is also
shown in the figure (i.e. line 3; black dotted line), where the permittivity and permeability of silver are respectively
$\varepsilon_1^{\rm metal}(\omega)=1-\omega_e^2/(\omega^2+i\gamma_e\omega)$ and
$\mu_1^{\rm metal}(\omega)=1$. Obviously,
the metal-dielectric interface has much larger Ohmic loss than the NIMM-dielectric interface.

Unfortunately, the suppression of the SP loss in the NIMM-dielectric interface is always accompanied by a de-confinement (or called de-localization)
of the SP because at $\omega=\omega_{\rm cri}$, Re$(1/k_2)$$\rightarrow \infty$; see line 2 in Fig.~\ref{FIG2} (the red dashed line)\,).  In order to acquire an acceptable suppression of the SP loss and a required SP confinement simultaneously, we are forced to select the signal-field frequency $\omega$ to have a small deviation from $\omega_{\rm cri}$~\cite{Kamli,Moi}. However, the deviation from $\omega_{\rm cri}$ will
make the electric field of the SP decay during propagation. For instance, if taking
$\omega=\omega_s=2\pi\times3.84\times10^{14}\,{{\rm s}^{-1}}$, one has
${\rm Im}(\beta)|_{\omega=\omega_s}=32.97\,\,{\rm cm^{-1}}$ (see the large blue solid circle on the line 1
of Fig.~\ref{FIG2}). That is to say, although the electromagnetic field can have a tight confinement within a scale of $k_2^{-1}|_{\omega=\omega_s}\approx985\,{\rm nm}$, which is still superior to conventional slab dielectric waveguides (without NIMM), a small loss exists simultaneously. In particular, the loss will be significant
for a long-distance propagation, hindering practical applications of SPs.

Such difficulty can be overcame by using the quantum emitters doped in the NIMM-dielectric
interface and working in the ARG scheme. When $N_a\neq0$, the positive imaginary part in the propagation constant caused by the Ohmic loss inherent in the NIMM can be completely eliminated by the negative imaginary part of $\chi_{\rm a}^{(1)}$ contributed by the gain from the quantum emitters. As an example of our model, we choose $^{87}\rm Rb$ atoms as the quantum emitters with $|1\rangle=|5^2S_{1/2},F=2\rangle$, $|2\rangle=|5^2S_{1/2},F=1\rangle$, $|3\rangle=|5^2P_{3/2},F=2\rangle$.
The system parameters are given as $\Gamma_{13}=\Gamma_{23}=\Gamma_{3}/2=\pi\times6.07\,{\rm MHz}$,
$\Gamma_{21}=0.5\,{\rm kHz}$, $\Gamma_{12}=0.01\,{\rm kHz}$,
$|{\bf p}_{31}|=3.58\times10^{-27}\,{\rm C\cdot cm}$, $\Delta_3=2.7\times10^9\,{\rm s}^{-1}$, $\Delta_2=10^4\,{\rm s}^{-1}$, $N_a=9.18\times10^{13}\,{\rm cm}^{-3}$, and $\Omega_p=1.4\times10^7\,{\rm s}^{-1}$. We obtain Im($\beta_a)|_{\omega=\omega_s+\Delta \omega}$ =0 with $\Delta \omega=-10^4\,{\rm s}^{-1}$,
which means that the quantum emitters can indeed provide a gain to compensate for the Ohmic loss in the NIMM. Thus the contradiction between the confinement and the
suppression of the Ohmic loss is resolved satisfactorily.

\vspace{5mm}
\noindent\textbf{\large \sf Linear Lossless Airy SPs}.
We now explore the possibility to get lossless Airy SPs excited at the NIMM-dielectric interface doped with quantum
emitters. To obtain linear Airy SP solutions, we solve Eqs.~(\ref{ME for NIMM}) and (\ref{ME for dielectric1}) under BCs~(\ref{B.C.}) and (\ref{B.C.2}) by employing the asymptotic expansion (similar to that used in Ref.~\cite{Marini})
${\bf E}_{\alpha}({\bf r})=g{\bf E}_{\alpha}^{(1)}+
g^{3/2}{\bf E}_{\alpha}^{(2)}+g^2 {\bf E}_{\alpha}^{(3)}+\cdots$,
with $g=(\omega-\omega_{0})/\omega_{0}$ and ${\bf E}_{\alpha}^{(m)}=
(E_{\alpha x}^{(m)}, E_{\alpha y}^{(m)},E_{\alpha z}^{(m)})$ ($m=1,2,3,...$)
being functions of $x$, $y_1=\sqrt{g}y$, and $z_{2}=gz$.
Here $\omega_{0}=\omega_{s}+\delta\omega$, a particular frequency determining a pure real propagation constant $\beta_{0}=\beta_{a}|_{\omega_{0}}$, as discussed above.
To give a consistent expansion, we further assume $\varepsilon_{2}+\chi_a^{(1)}=\varepsilon_{20}+\varepsilon_{21}$, with
$\varepsilon_{20}=(\varepsilon_2+\chi_{a}^{(1)})|_{\omega=\omega_0}$,
and $\varepsilon_{21}=(\omega-\omega_0)\cdot\frac{\partial\chi^{(1)}_a}{\partial\omega}|_{\omega=\omega_0}$.

Substituting the above expansions into Eqs.~(\ref{ME for NIMM}) and
(\ref{ME for dielectric1}), we obtain the unified form
\begin{equation}\label{ME with emitters}
\left[
\begin{array}{ccc}
k_{\alpha}^2 & 0 & i\beta_0\frac{\partial}{\partial x}\\
0 & \frac{\partial^2}{\partial x^2}-k_{\alpha}^2 & 0\\
0 & 0 & \frac{\partial^2}{\partial x^2}-k_{\alpha}^2
\end{array}
\right]
\left[
\begin{array}{l}
E_{\alpha x}^{(m)}\\E_{\alpha y}^{(m)}\\E_{\alpha z}^{(m)}
\end{array}
\right]
=
\left[
\begin{array}{l}
T_{\alpha x}^{(m)}\\T_{\alpha y}^{(m)}\\T_{\alpha z}^{(m)}
\end{array}
\right]
\end{equation}
($m=1,2,\cdots$), which can be solved order by order. Here $k_{1}=(\beta_0^2-k_0^2\varepsilon_{1}\mu_{1})^{1/2}$ and $k_{2}=[\beta_0^2-k_0^2 \varepsilon_{20}\mu_{2}]^{1/2}$ (both valued at $\omega=\omega_0$). For saving space,
the explicit expressions of $T_{\alpha j}^{(m)}$ ($j=x,y,z$; $m=1,2,\cdots$) and the expansions for BCs~(\ref{B.C.}) and (\ref{B.C.2}) are omitted here.

At the first order ($m=1$), we get the TM mode solution of Eq.~(\ref{ME with emitters})
\begin{eqnarray}\label{EM1}
&& {\bf E}^{(1)}_1({\bf r})=
\left({\bf e}_z-{\bf e}_x\frac{i\beta_0}{k_1}\right)A(y_1,z_2) e^{k_1x},\\
&& {\bf E}^{(1)}_2({\bf r})=
\left({\bf e}_z+{\bf e}_x\frac{i\beta_0}{k_2}\right)A(y_1,z_2) e^{-k_2x},
\end{eqnarray}
which are similar to Eqs.~(\ref{EM01}) and (\ref{EM02}), but here $A$ is an envelope function of the slow variables $y_1$ and $z_2$.

Solving Eq.~(\ref{ME with emitters}) at the second order ($m=2$) gives the solution of the signal field
${\bf E}_{1}^{(2)}({\bf r})={\bf e}_yB(y_1,z_2)e^{k_{1}x}$,
${\bf E}_{2}^{(2)}({\bf r})={\bf e}_yB(y_1,z_2)e^{-k_{2}x}$,
where $B$ is another slowly-varying envelope function. The boundary condition for magnetic field ${\bf H}_{\alpha}
({\bf r})$, i.e.  ${\bf e}_x\times({\bf H}_2({\bf r})-{\bf H}_1({\bf r})\,)=0$ with ${\bf H}_{\alpha}({\bf r})=[1/(i\omega\mu_0\mu_{\alpha})] \nabla\times {\bf E}_{\alpha}({\bf r})$ ($\alpha=1,2$), at this order
yields the relation $B=-(i/\beta_0)\partial A/\partial y_1$. We see that to this order
the signal field is no longer a TM wave since a $y$-component of the electric field appears.

Similarly, solving Eq.~(\ref{ME with emitters}) at the third order ($m=3$)
we obtain the solution of the signal field, which is presented in Method.
The BC at this order results in
$2i\beta_0 \partial A/\partial z_2+\partial^2 A/\partial y_1^2+f\,A=0$,
where $f=\varepsilon_{21}D_0$ with $k_{0}=\omega_{0}/c$, and $D_0=k_0^2 (\mu_1k_1+\mu_2k_2)k_1^2(\beta_0^2+k_2^2)/[k_2(k_1^2-k_2^2)^2]$. Returning to original
variables and making the transformation $y=R_y\eta$, $z=L_{\rm Diff}s$, and $A=U_0 u\exp[ifz/(2\beta_0)]$,
with $R_{y}$, $L_{\rm Diff}\equiv\beta_0 R_{y}^2$, and $U_0$ being, respectively,
typical beam radius, diffraction length, and amplitude of the signal field, we obtain the dimensionless equation
\begin{equation}\label{NDLSE}
i\frac{\partial u}{\partial s}+\frac{1}{2}\frac{\partial^2 u}{\partial \eta^2}=0,
\end{equation}
which admits the Airy beam solution~\cite{Berry}
\begin{equation}
u(\eta,s)=V_0 {\rm Ai}\left(\eta-\frac{s^2}{4}\right) \,e^{ i\frac{s}{2}\left(\eta-\frac{s^2}{6}\right) },
\label{Airy dimensionless solution0}
\end{equation}
with $V_{0}^{-1}={\rm max}[{\rm Ai}(\eta)]$, a constant introduced for making the peak intensity
of $u$ to be 1.

Using the above expressions, we obtain the explicit expression of ${\bf E}$ for the Airy SP
propagating down to the NIMM-dielectric interface

\begin{eqnarray}
&& {\bf E}_1({\bf r},t;\omega_0)=\left({\bf e}_z-{\bf e}_x\frac{i\beta_0}{k_1}-{\bf e}_y\frac{i}{\beta_0}\frac{\partial}{\partial y}\right)A(y,z)e^{i\left(\beta_0z-\omega_0 t\right)}e^{k_1 x}+{\rm c.c.},\label{generic electric expression1}\\
&& {\bf E}_2({\bf r},t;\omega_0)=\left({\bf e}_z+{\bf e}_x\frac{i\beta_0}{k_2}-{\bf e}_y\frac{i}{\beta_0}\frac{\partial}{\partial y}\right)A(y,z)e^{i\left(\beta_0z-\omega_0 t\right)}e^{-k_2 x}+{\rm c.c.},\label{generic electric expression2}
\end{eqnarray}
with $A(y,z)=U_0 V_0 {\rm Ai} \left( \frac{y}{R_y}-\frac{z^2}{ 4L_{\rm Diff}^2 } \right)
e^{i\frac{z}{2L_{\rm Diff}} \left(\frac{y}{R_y}-\frac{z^2}{6L_{\rm Diff}}\right) }
e^{i\frac{f}{2\beta_0}z}$.

The Airy SP solution~(\ref{generic electric expression1}) and~(\ref{generic electric expression2}) has three notable features.
(i)~It is non-diffractive and bends along the parabolic trajectory $y=z^2/(4\beta_0^2R_y^3)$.
(ii)~Generally, $f$ is a complex number when $\omega \neq \omega_{0}$, which means that the amplitude
of the Airy SP increases or decreases during propagation. However at $\omega=\omega_{0}$ one has
$f= 0$ and hence the solution~(\ref{generic electric expression1}) and~(\ref{generic electric expression2}) has no attenuation upon propagation.
We call such solution as {\it lossless Airy SP}. The reasons for the lossless propagation of the Airy SP are due to the contributions by the destructive interference effect between the electric and magnetic responses of the NIMM and by the ARG from the quantum emitters.

However, the Airy function solution (\ref{Airy dimensionless solution0}) is of infinite energy,
which is not realistic and unobservable. A finite energy (or truncated) Airy beam solution
of Eq.~(\ref{NDLSE}) is given by~\cite{Siv1}
$u(\eta,s)=V_a{\rm Ai}\left(\eta-\frac{s^2}{4}+ias\right)e^{i\frac{s}{2}\left(\eta-\frac{s^2}{6}\right)}
e^{a\eta-\frac{as^2}{2}+i\frac{a^2s}{2}}$,
where $a$ being a small, positive real number (apodization parameter) introduced to make the ideal Airy beam
have an finite energy, and $V_a^{-1}\equiv {\rm max}[{\rm Ai}(\eta)\exp(a\eta)]$ is an auxiliary
factor introduced for convenience (i.e. for setting the beam peak intensity to be 1 for any $a$).
Then the explicit expression of electric field for the finite energy Airy SP upon propagation
is still given by Eqs.~(\ref{generic electric expression1}) and~(\ref{generic electric expression2}) but with
\begin{equation}
A(y,z)=U_0 V_a{\rm Ai}\left(\frac{y}{R_y}-\frac{z^2}{4L_{\rm Diff}^2}+ia\frac{z}{L_{\rm Diff}}\right)
e^{i\frac{z}{2L_{\rm Diff}}\left(\frac{y}{R_y}-\frac{z^2}{6L_{\rm Diff}}\right)}e^{a\frac{y}{R_y}-\frac{az^2}{2L_{\rm Diff}^2}+i\frac{a^2z}{2L_{\rm Diff}}}e^{i\frac{f}{2\beta_0}z}.
\label{Airy dimension solution}
\end{equation}

\begin{figure}
\includegraphics[width=\columnwidth]{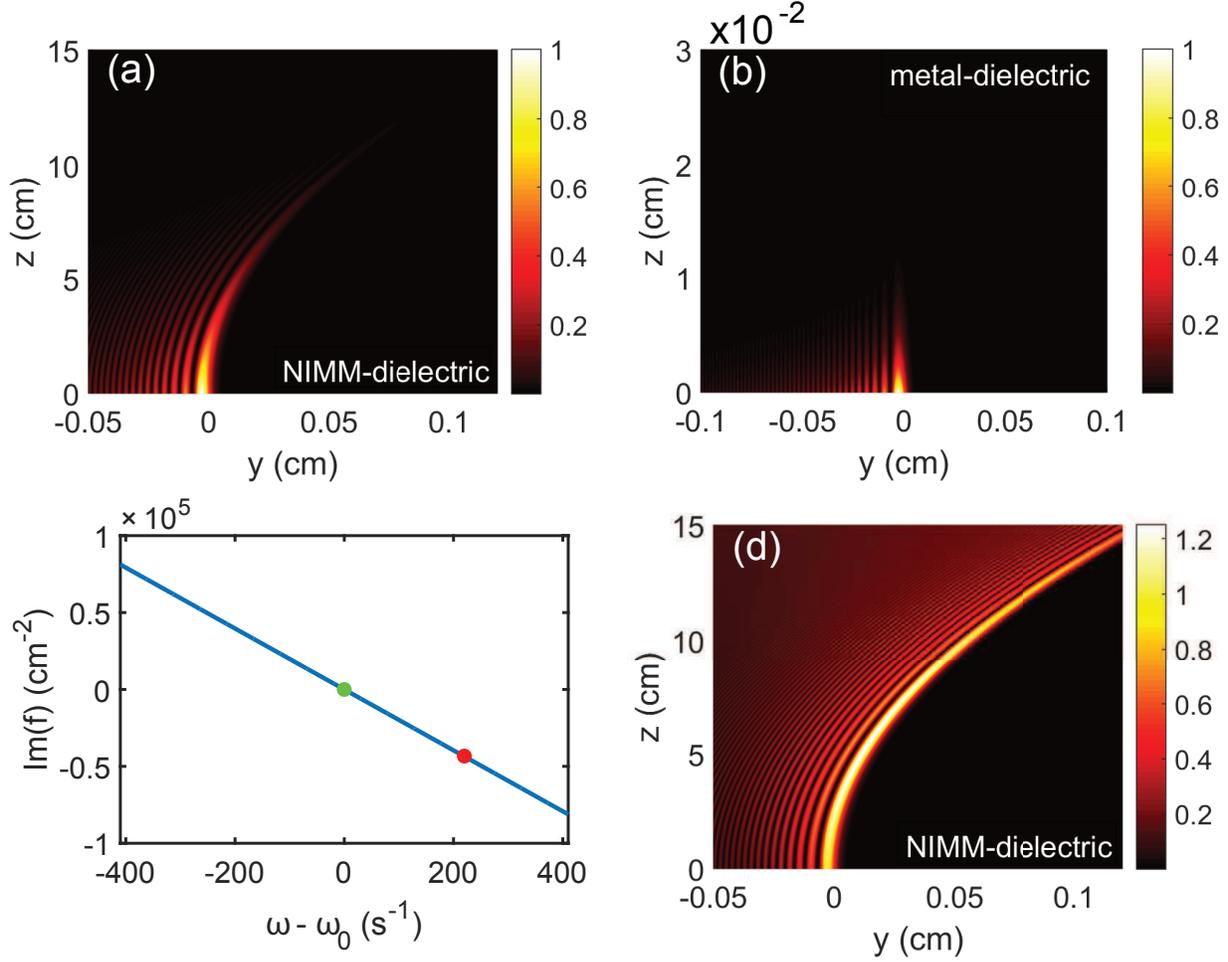}
\caption{\protect\footnotesize {\bf Linear lossless Airy SPs.} (a) $|u|^2$ of the finite energy Airy SP propagating
upon the NIMM-dielectric interface (i.e. the $y$-$z$ plane) for $\omega=\omega_{0}$. (b) $|u|^2$ of the finite energy Airy SP propagating upon a metal-dielectric interface for $\omega=\omega_{0}$. (c) ${\rm Im}(f)$ for the NIMM-dielectric interface as a function of $\omega-\omega_{0}$, the green (red) solid circle indicates the particular value of Im($f$) at $\omega-\omega_{0}=0$ ($\omega-\omega_{0}=220\,{\rm s}^{-1}$). (d)
$|u|^2$ of the finite energy Airy SP propagating along the NIMM-dielctric interface for $\omega-\omega_{0}=220\,{\rm s}^{-1}$ (corresponding the red solid circle in panel (c)\,).}
\label{Fig3}
\end{figure}

Shown in Fig.~\ref{Fig3}(a)
is the intensity profile $|u|^2$ of the finite energy Airy SP propagating along the NIMM-dielectric interface
(i.e. the $y$-$z$ plane) for $\omega=\omega_{0}$.
When plotting the figure, we have chosen $a=0.02$ and $R_y=30\,{\rm \mu m}$ and the other system parameters the same as used above.
Fig.~\ref{Fig3}(b) shows the propagation of the intensity profile $|u|^2$ of the finite energy Airy
SP in a similar system where the NIMM is replaced by a metal (i.e. a metal-dielectric interface).
We see that the Airy SP undergoes no obvious absorption when propagating upon the NIMM-dielectric interface.
In contrast, the Airy SP propagating along
the metal-dielectric interface has a significant propagation loss and hence it attenuates very rapidly
so that the bending of its motional trajectory cannot even be observed (Fig.~\ref{Fig3}(b)\,).

The system may acquire a neat gain through the quantum emitters. Fig.~\ref{Fig3}(c) shows  Im($f$) for the NIMM-dielectric interface. For $\omega=\omega_{0}$, ${\rm Im}(f)=0$ (the green solid circle in Fig.~\ref{Fig3}(c)\,), and hence the system has a exact balance between loss and gain and a lossless Airy SP can be excited.
For $\omega>\omega_{0}$, ${\rm Im}(f)<0$ (the red solid circle in Fig.~\ref{Fig3}(c)\,), and
hence the system has a neat gain, which can be used to incompletely compensate the Ohmic loss in the
NIMM and also the loss resulted by the introduction of the positive apodization parameter $a$.
In this case, the Airy SP can propagate to a long distance without any attenuation.
Fig.~\ref{Fig3}(d) shows the intensity profile $|u|^2$ of the Airy SP for $\omega-\omega_{0}=220\,{\rm s}^{-1}$.
We see that, comparing with Fig.~\ref{Fig3}(a), instead of attenuation the Airy SP has indeed
a gain during propagation.

\vspace{5mm}
\noindent\textbf{\large \sf Airy surface polaritonic solitons}{\label{sec3}}. Because
for $\omega>\omega_{0}$ the system has a neat gain,
the Airy SP will be amplified when propagating along the NIMM-dielectric interface.
For a long propagation the Airy SP will be amplified significantly,
the linear theory given above is no longer valid. Thus it is necessary
to extend the linear theory to a nonlinear regime and consider the possibility to
generate lossless Airy surface polaritonic solitons in the system.

To this end, we assume that the nonlinear effect in the system comes only from the quantum emitters
due to the resonant character of the interaction between the EM field with the quantum emitters. To derive a
envelope equation for the signal field with a weak nonlinearity, we assume the perturbation expansion
${\bf E}_{\alpha}({\bf r})=g^{1/2}{\bf E}_{\alpha}^{(1)}+
g{\bf E}_{\alpha}^{(2)}+g^{3/2}{\bf E}_{\alpha}^{(3)}+\cdots$ ($\alpha=1,2$),
with ${\bf E}_{\alpha}^{(m)}$ ($m=1,2,3,...$) the functions of the multi-scale variables
$x$, $y_1=\sqrt{g}y$, and $z_{2}=gz$. Substituting this expansion into Eqs.~(\ref{ME for NIMM}) and
(\ref{ME for dielectric1}), we obtain a set of equations similar to those given in Eq.~(\ref{ME with emitters}),
which can be solved order by order.

At the first two orders ($m=1,2$), we obtain solutions of the signal field,
which are the same as those given in the linear case presented above.
The solution at the third order ($m=3$) is given in Method.
The BC of at this order is nonlinear, which results in
the equation for the envelope $A$ as
$2i\beta_0 \partial A/\partial z_2+\partial^2 A/\partial y_1^2+f\,A+W\,|A|^2 A=0$. Here
$W=k_{0}^2 W_{\rm m}|_{\omega=\omega_{0}}\chi_{a0}^{(3)}$ is nonlinear coefficient, with $k_{0}=\omega_{0}/c$, $W_{\rm m}|_{\omega=\omega_{0}}=k_1^2(\mu_1k_{1}+\mu_2k_{2})(\beta_{0}^2+|k_2|^2)(\beta_{0}^2+k_2^2)/\{(k_2^2-k_1^2)^2|k_2|^2[{\rm Re}(k_2)+k_2]\}$ and $\chi^{(3)}_{a0}=\chi^{(3)}_{a}|_{\omega=\omega_{0}}$. The real part of $W$ (i.e.
${\rm Re}(W)$) accounts for the self-phase modulation (SPM) effect corresponding to the self-focusing (for ${\rm Re}(W)>0$) or self-defocusing (for ${\rm Re}(W)<0$). Here we  focus only on the self-focusing in order to
generate bright Airy surface polaritonic solitons.
After returning to the original variables and making the transformation $y=R_y\eta$, $z=L_{\rm Diff}s$,
and $A=U_0 u$, the above equation convertes into the dimensionless form
\begin{equation}\label{NDGLE}
i\frac{\partial u}{\partial s}+\frac{1}{2}\frac{\partial^2 u}{\partial \eta^2}+l_{\rm r}\,u+g_{\rm r}|u|^2\,u=-i\left[l_{\rm i}\,u+g_{\rm i}|u|^2\,u\right],
\end{equation}
with $l_{\rm r}=L_{\rm diff}{\rm Re}(f)/(2\beta_{0})$, $l_{\rm i}=L_{\rm diff}{\rm Im}(f)/(2\beta_{0})$ and
$g_{\rm r}={\rm Sgn[{\rm Re}(W)]}L_{\rm diff}/L_{\rm NL}$, $g_{\rm i}=U_0^2L_{\rm diff}{\rm Im}(W)/(2\beta_{0})$,
where $L_{\rm NL}=2\beta_0/\left[U_0^2|{\rm Re}(W)|\right]$ is typical nonlinearity length. For obtaining a
stable soliton, one requires a balance between the diffraction and the nonlinearity, i.e. $L_{\rm diff}=L_{\rm NL}$,
and thus $U_0=\sqrt{2/[R_y^2|{\rm Re}(W)|]}$.

Although Eq.~(\ref{NDGLE}) has complex coefficients, the imaginary parts of these coefficients
can be made much smaller than their real parts due to the contribution by the ARG induced by the quantum emitters,
and hence one can generate Airy surface polaritonic solitons when the initial profile of the signal-field envelope is an Airy function. For this aim, we give a realistic parameter set for the formation of an Airy surface polaritonic soliton in the system. By selecting $R_y=30\,{\rm \mu m}$,  $\omega-\omega_{0}=400\,{\rm s}^{-1}$, and other parameters the same as those given in above discussion,
we thus obtain $L_{\rm diff}=L_{\rm NL}=1.14\,{\rm cm}$, $U_0=1.23\times10^{-2}\,{\rm V/cm}$, and the dimensionless coefficients of the equations $l_{\rm r}=-4.59$, $l_{\rm i}=-0.35$, $g_r=1$, and $g_{\rm i}=-0.16\times10^{-2}$. One can see that the imaginary part of the coefficients are indeed much smaller than their corresponding real parts, hence in the leading order the terms on the right side of the Eq.~(\ref{NDGLE}) can be safely
neglected.
\begin{figure}
\includegraphics[width=\columnwidth]{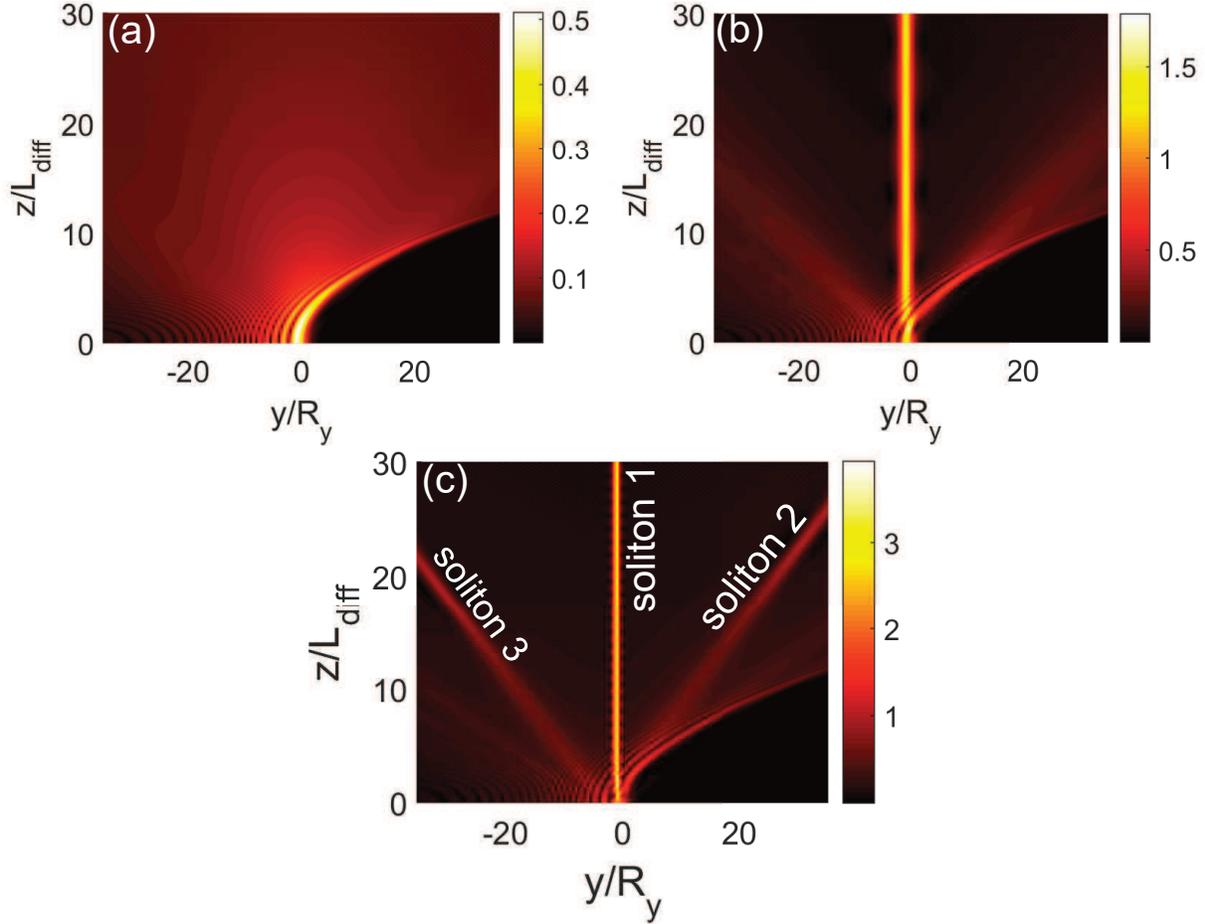}
\caption{\protect\footnotesize{\bf Airy surface polaritonic solitons.}  Nonlinear evolution of $|u|$ as functions of $y/R_y$ and $z/L_{\rm Diff}$ for different $u_0$. (a) $u_0=0.5$: Airy beam with shed CW radiations; (b) $u_0=1.3$: Airy beam with shed
static surface polaritonic soliton (i.e. the straight bright strip near at $y=0$) and CW radiations.
(c) $u_0=2.4$: Airy beam with shed static surface polaritonic soliton (i.e. ``soliton 1''), the pair of moving surface polaritonic solitons (i.e. ``soliton 2'' and ``soliton 3''), and CW radiations. }
\label{Fig4}
\end{figure}
With the SPM coefficient $W$ given above it is easy to estimate the optical Kerr effect
of the system by using the formulas
$n=n_0+n_2 I$ and $n_2=W/(2\beta_0^2)$, where $n$ is total refractive index,
$n_0$ is linear refractive index, $n_2$ is Kerr coefficient, and
$I$ is the light intensity of the signal field. Based on the above parameters we obtain
$n_2=\left(4.58-i0.73\right)\times10^{-2}$ cm$^2$/V$^{2}$, which is quite large comparing with
conventional systems (such as optical fibers).

We numerically solve Eq.~(\ref{NDGLE}) by using a split-step Fourier method, with the initial condition given by
$u(s=0,\eta)=u_0\,V_{a}{\rm Ai}(\eta)e^{a\eta}$,
where $u_0$ is an amplitude parameter, and $V_{a}^{-1}\equiv{\rm max}[{\rm Ai}(\eta)\exp(a\eta)]$ is, as defined above, a normalization factor of the amplitude dependent on the apodization factor $a$.
Fig.~\ref{Fig4} shows the evolution of $|u|$
as a function of $y/R_y$ and $z/L_{\rm Diff}$ for different $u_0$, with $a=0.06$.
We see that for a smaller $u_0$ (i.e. $u_0=0.5$)  the Airy beam has a shedding of CW radiations  (Fig.~\ref{Fig4}(a)\,) during propagation.  However, as $u_0$ increases (i.e. $u_0=1.3$), a static surface polaritonic soliton (i.e. the straight bright strip near at $y=0$) is shed from the Airy beam (Fig.~\ref{Fig4}(b)\,),
with additional CW radiations. As $u_0$ increases further (i.e. $u_0=2.4$), besides the appearance of a
static surface polaritonic soliton (``soliton 1'' in Fig.~\ref{Fig4}(c)\,) which displays an obvious oscillation
along $z$-axis, a pair of moving surface polaritonic solitons (i.e. ``soliton 2'' and ``soliton 3''
in Fig.~\ref{Fig4}(c)\,) is also generated. Two solitons in the pair have the same amplitude and opposite velocity,
ensuring the conservation of the total momentum in the system. In this case, except for the production of the static
soliton and the moving soliton pair, some CW radiations are also appear.
Although these phenomena are similar to those
found in Refs.~\cite{Fattal,Tsoy}, what we explored here is for Airy surface polaritonic solitons, which are not reported
in literature up to now.

The threshold of the optical power density for generating the Airy surface polaritonic solitons can be calculated
by using Poynting's vector~\cite{Huang1}, which reads ${\bar P}_{\rm max}=1.47\,{\rm nW}$. Thus for generating
the Airy surface polaritonic solitons very low input power is needed.
%

\vspace{6mm}
\noindent\textbf{\Large \sf Discussion}

The analysis presented above showed that lossless propagation of linear and nonlinear Airy SPs can be realized indeed via ARG. We now  make some remarks on them. First, we have assumed, like that done in Refs.~\cite{Kamli,Moi,Kamli1,Sio}, the NIMM is spatially homogeneous. Such assumption requires the lattice constant $d$ of the array of meta-atoms (i.e. artificial subwavelength building blocks) in the NIMM must be at least one order of magnitude smaller than the wavelength $\lambda_s$ of the signal field. In this situation, the NIMM can be taken as an effective and spatially homogeneous medium. In our model, $\lambda_s=780$ nm (i.e. at the red end of visible spectrum), hence $d$ must be less than 100 nm. Such optical NIMMs may be designed by using double-fishnet structures and are now available experimentally (see Refs.~\cite{Wue,ZZhang,ZhangShuang,Shalaev1,Dolling}). However, in NIMMs there exist inhomogeneities due to the roughness of sample and  the fluctuations in the meta-atom size of the meta-atom array, which may result in an inhomogeneous broadening for the absorption spectrum of the signal field. Related calculation including such inhomogeneous broadening can be carried out in our theoretical scheme, which is, however, beyond the scope of the present work.

Second, in our analysis the influences from the lower boundary of the NIMM and the upper boundary of the dielectric have been neglected. Such assumption is valid when the thicknesses of both the NIMM and the dielectric are large enough. Additionally, because both the linear and nonlinear Airy SP beams have curved trajectories during propagation, the theoretical approach presented above maybe violate the assumption of paraxial approximation~\cite{Sala}. By simple calculations based on the results in Fig.~\ref{Fig3} and Fig.~\ref{Fig4}, we obtain the deflection angles of the linear and nonlinear Airy SPs to be about $2.3\times10^{-3}\,{\rm rad}$ and $2.6\times10^{-3}\,{\rm rad}$, respectively. Such small deflection angles ensure the validity of the paraxial approximation used in the derivation of the envelope equations (\ref{NDLSE}) and (\ref{NDGLE}). In the case of large deflection angle, the paraxial approximation is broken and hence the approach given above must be generalized~\cite{ZhangP1,ZhangP2,ZhangP3}.

Third, noise is usually an important problem for a system where SPs are compensated by a gain medium. In our analysis the noise problem
is not considered since it is another topic beyond the scope of the present work. We should, however, point out that
in our system the noise induced by the ARG gain is not significant because in our consideration the one-photon detuning $\Delta_3$ is taken to be large (order of GHz), and hence the population in the level $|3\rangle$ is very small ($\sigma_{33}^{(0)}\approx 2.40\times10^{-5}$).
As a result, the gain contributed by the quantum emitters is not large and thus the noise induced by the gain,
attributing to amplified spontaneous emission, plays a negligible role.
Note that a significant gain is not needed in our scheme because the Ohmic loss in the NIMM has
already been greatly suppressed by the destructive interference effect between the electric and magnetic responses in the NIMM; see Fig.~\ref{FIG2} and related discussions. In addition, in our system the photon number in the signal field is large ($\approx$ 5800), and hence the noise induced by the quantum effect of the signal field can be neglected.

In summary, we have proposed a scheme for realizing a lossless propagation of linear and nonlinear Airy surface
polaritons in a NIMM-dielectric interface where three-level quantum emitters working in an ARG regime are doped.
By using the ARG from the quantum emitters and the destructive interference effect between the electric and magnetic
responses from the NIMM, we have shown that not only the Ohmic loss of the NIMM but also the light absorption
of the quantum emitters can be completely eliminated. As a result, non-diffractive Airy SPs can propagate
for a very long distance without attenuation. We have also shown that the Kerr nonlinearity of the system can be
largely enhanced due to the contribution of the quantum emitters, and hence lossless Airy surface polaritonic
solitons propagating down the NIMM-dielectric interface with very low power can be realized in the system.
The lossless Airy SPs predicted here may have not only fundamental interest in the research of nanophotonics
but also promising applications for the light information processing and transmission by using active
micro-nano structures.

\vspace{5mm}
\noindent\textbf{\Large \sf Methods}\\
{\small

\noindent\textbf{\large \sf Bloch equations and solution for $\sigma_{32}$}.
Explicit expression of the Bloch equation describing the motion of the quantum emitters with the three-level ARG
configuration reads

\begin{eqnarray}
&& i\left(\frac{\partial}{\partial t}+\Gamma_{21}\right)\sigma_{11}-i\Gamma_{12}\sigma_{22}-i\Gamma_{13}\sigma_{33}-\Omega_{p}\sigma_{31}^{\ast}+\Omega_{p}^{\ast}\sigma_{31}=0,\label{BEstart}\\
&& i\left(\frac{\partial}{\partial t}+\Gamma_{12}\right)\sigma_{22}-i\Gamma_{21}\sigma_{11}-i\Gamma_{23}\sigma_{33}-\Omega_{s}({\bf r})\sigma_{32}^{\ast}
+\Omega_{s}^{\ast}({\bf r})\sigma_{32}=0,\\
&& i\left(\frac{\partial}{\partial t}+\Gamma_{13}+\Gamma_{23}\right)\sigma_{33}+\Omega_{p}\sigma_{31}^{\ast}-\Omega_{p}^{\ast}\sigma_{31}+\Omega_{s}({\bf r})\sigma_{32}^{\ast}-\Omega_{s}^{\ast}({\bf r})\sigma_{32}=0,\\
&& \left(i\frac{\partial}{\partial t}+d_{21}\right)\sigma_{21}+\Omega_s^{\ast}({\bf r})\sigma_{31}-\Omega_p\sigma_{32}^{\ast}=0,\\
&& \left(i\frac{\partial}{\partial t}+d_{31}\right)\sigma_{31}+\Omega_p(\sigma_{11}-\sigma_{33})+\Omega_s({\bf r})\sigma_{21}=0,\\
&& \left(i\frac{\partial}{\partial t}+d_{32}\right)\sigma_{32}+\Omega_s({\bf r})(\sigma_{22}-\sigma_{33})+\Omega_p\sigma_{21}^{\ast}=0,
\label{BEend}
\end{eqnarray}
where $\Omega_{s}({\bf r})=({\bf E}_{2}({\bf r})\exp(i\beta_a z)\cdot{\bf p}_{32})/\hbar$\,\, ($\Omega_{p}=({\bf E}_p\cdot{\bf p}_{31})/\hbar$) is the half Rabi frequency of the signal (pump) field, with ${\bf E}_p$ being the electric field of the pump laser (which is a constant vector and given). $d_{21}=\Delta_2+i\gamma_{21}$, $d_{31}=\Delta_3+i\gamma_{31}$, and $d_{32}=\Delta_3-\Delta_2+i\gamma_{32}$. Here
$\gamma_{jl}=(\Gamma_{j}+\Gamma_{l})/2+\gamma_{jl}^{\rm col}$ are coherence decay rates ($\Gamma_{l}\equiv\sum_{E_{j}<E_{l}}\Gamma_{jl}$), with $\Gamma_{jl}$ denoting the population decay rate and $\gamma_{jl}^{\rm col}$ denoting the dipole dephasing rate from the state $|l\rangle$ to the state $|j\rangle$.

\noindent\textbf{\large \sf Solution of $\sigma_{32}$}.
We first give some remarks on the Eqs.~(\ref{BEstart}-\ref{BEend}). (i)~In our ARG excitation scheme, the detuning $\Delta_3$
is assumed to be large enough so that inhomogeneous (energy-level) broadening of the emitters can be largely
suppressed. (ii)~As usual~\cite{Huang1}, the pump field is taken to be strong enough so that its
depletion is negligible (i.e. $\Omega_p$ is a constant) during the propagation of the signal field.
(iii)~For CW excitations the time derivatives in Eqs.~(\ref{BEstart}-\ref{BEend}) can be safely neglected because the time duration $\tau_{0}$ of the pulsed signal field satisfies the condition $\tau_0\gamma_{\rm max}\gg 1$, with $\gamma_{\rm max}$ being the maximum decay rate of the quantum emitters in the system. Therefore one can get $\sigma_{32}$ by solving Eqs.~(\ref{BEstart}-\ref{BEend}) algebraically.

For solving $\sigma_{32}$, we make the expansion
$\Omega_{s}=\Omega_{s}^{(1)}+\Omega_{s}^{(2)}+\Omega_{s}^{(3)}+\cdots$,
$\sigma_{jk}=\sigma_{jk}^{(0)}+\sigma_{jk}^{(1)}+\sigma_{jk}^{(2)}+\sigma_{jk}^{(3)}+\cdots$. Then
Eqs.~(\ref{BEstart}-\ref{BEend}) can be solved order by order. Base state solution reads
\begin{eqnarray}\label{base state}
&& \sigma_{11}^{(0)}=\frac{\Gamma_{12}(\Gamma_{13}+\Gamma_{23}+D)}{(\Gamma_{12}+\Gamma_{21})(\Gamma_{13}
+\Gamma_{23})+(2\Gamma_{12}+\Gamma_{21}+\Gamma_{23})D},\\
&&\sigma_{22}^{(0)}=\frac{\Gamma_{21}(\Gamma_{13}+\Gamma_{23})+(\Gamma_{21}+\Gamma_{23})D}{(\Gamma_{12}
+\Gamma_{21})(\Gamma_{13}+\Gamma_{23})+(2\Gamma_{12}+\Gamma_{21}+\Gamma_{23})D},\\
&& \sigma_{33}^{(0)}=\frac{\Gamma_{12}D}{(\Gamma_{12}+\Gamma_{21})(\Gamma_{13}+\Gamma_{23})+(2\Gamma_{12}
+\Gamma_{21}+\Gamma_{23})D},\\
&& \sigma_{31}^{(0)}=-\frac{\Omega_p}{d_{31}}\frac{\Gamma_{12}(\Gamma_{13}+\Gamma_{23})}{(\Gamma_{12}
+\Gamma_{21})(\Gamma_{13}+\Gamma_{23})+(2\Gamma_{12}+\Gamma_{21}+\Gamma_{23})D},
\end{eqnarray}
and $\sigma_{32}^{(0)}=\sigma_{21}^{(0)}=0$, with $D=2\gamma_{31}|\Omega_p|^2/|d_{31}|^2$.
For large $\Delta_3$, the above expressions are reduced to $\sigma_{11}^{(0)}\approx \Gamma_{12}/(\Gamma_{12}+\Gamma_{21})$, $\sigma_{22}^{(0)}\approx \Gamma_{21}/(\Gamma_{12}+\Gamma_{21})$, $\sigma_{33}^{(0)}\approx0$, and $\sigma_{31}^{(0)}\approx -\Omega_p\Gamma_{12}/[d_{31}(\Gamma_{12}+\Gamma_{21})]$.

The solution of the first order reads $\sigma_{32}^{(1)}=a_{32}^{(1)}\Omega_{s}^{(1)}$, $\sigma_{21}^{(1)\ast}=a_{21}^{(1)}\Omega_{s}^{(1)}$, and other $\sigma_{jk}^{(1)}=0$, where
\begin{eqnarray}
&& a_{32}^{(1)}=\frac{\Omega_p\sigma_{31}^{(0)\ast}+d_{21}^{\ast}\left(\sigma_{33}^{(0)}-\sigma_{22}^{(0)}\right)}{|\Omega_p|^2+d_{21}^{\ast}d_{32}},\\
&& a_{21}^{(1)}=-\frac{d_{32}\sigma_{31}^{(0)\ast}-\Omega_p^{\ast}\left(\sigma_{33}^{(0)}-\sigma_{22}^{(0)}\right)}{|\Omega_p|^2+d_{21}^{\ast}d_{32}}.
\end{eqnarray}

The solution of the second order reads $\sigma_{jj}^{(2)}=a_{jj}^{(2)}|\Omega_{s}^{(1)}|^2$ ($j=1,2,3$), $\sigma_{31}^{(2)}=a_{31}^{(2)}|\Omega_{s}^{(1)}|^2$, and other $\sigma_{jk}^{(2)}=0$, where
\begin{eqnarray}
&& a_{31}^{(2)}=-\frac{1}{d_{31}}\left[a_{21}^{(1)\ast}+\Omega_p\left(2a_{11}^{(2)}+a_{22}^{(2)}\right)\right],\\
&& a_{11}^{(2)}=i\frac{\left(\Gamma_{12}+\Gamma_{23}\right)\left(\frac{\Omega_p}{d_{31}^{\ast}}a_{21}^{(1)}-\frac{\Omega_p^{\ast}}{d_{31}}a_{21}^{(1)\ast}\right)+\left(-\Gamma_{12}+\Gamma_{13}+D\right)\left(a_{32}^{(1)\ast}-a_{32}^{(1)}\right)}{(\Gamma_{12}
+\Gamma_{21})(\Gamma_{13}+\Gamma_{23})+(2\Gamma_{12}+\Gamma_{21}+\Gamma_{23})D},\\
&& a_{22}^{(2)}=i\frac{\left(\Gamma_{21}-\Gamma_{23}\right)\left(\frac{\Omega_p}{d_{31}^{\ast}}a_{21}^{(1)}-\frac{\Omega_p^{\ast}}{d_{31}}a_{21}^{(1)\ast}\right)-\left(\Gamma_{21}+\Gamma_{13}+2D\right)\left(a_{32}^{(1)\ast}-a_{32}^{(1)}\right)}{(\Gamma_{12}
+\Gamma_{21})(\Gamma_{13}+\Gamma_{23})+(2\Gamma_{12}+\Gamma_{21}+\Gamma_{23})D},\\
&& a_{33}^{(2)}=-(a_{11}^{(2)}+a_{22}^{(2)}),
\end{eqnarray}

The solution of the third order reads $\sigma_{32}^{(3)}=a_{32}^{(3)}|\Omega_{s}^{(1)}|^2\Omega_{s}^{(1)}$, where
\begin{equation}
a_{32}^{(3)}=\frac{\Omega_pa_{31}^{(2)\ast}+d_{21}^{\ast}\left(a_{33}^{(2)}-a_{22}^{(2)}\right)}{|\Omega_p|^2+d_{21}^{\ast}d_{32}},
\end{equation}
and others are omitted here.

with the above results and using the definition $\Omega_{s}({\bf r})=({\bf E}_{2}({\bf r})\exp(i\beta_a z)\cdot{\bf p}_{32})/\hbar$, we obtain
[up to the third order in ${\bf E}_{2}({\bf r})$]
\begin{equation}\label{psigma32}
{\bf p}_{23}\sigma_{32}=\frac{|{\bf p}_{23}|^2}{\hbar}a_{32}^{(1)} {\bf E}_2({\bf r})e^{i\beta_a z}
+\frac{|{\bf p}_{23}|^4}{\hbar^3}a_{32}^{(3)} |{\bf E}_2({\bf r})|^2e^{-2{\rm Im}(\beta_{\rm a})z} {\bf E}_2({\bf r})e^{i\beta_a z}.
\end{equation}

\vspace{4mm}

\noindent\textbf{\large \sf Expressions of the signal field at the third order ($m=3$)}.
In the linear case, the expression of the solution of the signal field at the third order is given by
\begin{eqnarray}
&& {E}^{(3)}_{1z}=\left[A^{(3)}_{1}+\frac{L_{1z}}{2k_1}x\right]e^{k_1x},\label{Ez1}\\
&& {E}^{(3)}_{1x}=\left[-\frac{i\beta_0}{k_1}A^{(3)}_1+\frac{i\beta_0}{2k_1^3}L_{1z}-\frac{i\beta_0}{2k_1^2}L_{1z}x+\frac{L_{1x}}{k_1^2}\right]e^{k_1x},\label{Ex1}\\
&& {E}^{(3)}_{2z}=\left[A^{(3)}_{2}-\frac{L_{2z}}{2k_2}x\right]e^{-k_2x},\\
&& {E}^{(3)}_{2x}=\left[\frac{i\beta_0}{k_2}A^{(3)}_2-\frac{i\beta_0}{2k_2^3}L_{2z}-\frac{i\beta_0}{2k_2^2}L_{2z}x+\frac{L_{2x}}{k_2^2}\right]e^{-k_2x},
\end{eqnarray}
where $A^{(3)}_{\alpha}$ ($\alpha=1,2$) are constants, and
\begin{eqnarray}
&& L_{1x}=\frac{\beta_0^2+\varepsilon_{10}\mu_1k_0^2}{k_1}\frac{\partial A}{\partial z_2}-{k_1}\frac{\partial B}{\partial y_1}-\frac{i\beta_0}{k_1}\frac{\partial^2A}{\partial y_1^2},\\
&& L_{2x}=-\frac{\beta_0^2+\varepsilon_{20}\mu_2k_0^2}{k_2}\frac{\partial A}{\partial z_2}+{k_2}\frac{\partial B}{\partial y_1}+\frac{i\beta_0}{k_2}\frac{\partial^2A}{\partial y_1^2}+\frac{i\beta_0}{k_2}k_0^2\mu_2\varepsilon_{21}A,\\
&& L_{1z}=-2i\beta_0\frac{\partial A}{\partial z_2}-\frac{\partial^2A}{\partial y_1^2},\\
&& L_{2z}=-2i\beta_0\frac{\partial A}{\partial z_2}-\frac{\partial^2A}{\partial y_1^2}-k_0^2\mu_2\varepsilon_{21}A.
\end{eqnarray}
The expressions of $E_{\alpha y}^{(3)}$ are not needed and thus omitted here.

In the nonlinear case, the expressions of $E_{1j}^{(3)}$  $(j=x,z)$ at the third order are the same as those in the linear case given above, while the expressions of $E_{2j}^{(3)}$  $(j=x,z)$ read
\begin{eqnarray}
&& {E}^{(3)}_{2z}=\left[A^{(3)}_{2}-\frac{L_{2z}}{2k_2}x\right]e^{-k_2x}+\frac{N_{z}}{4{\rm Re}(k_2)\left[{\rm Re}(k_2)+k_2\right]}e^{-2{\rm Re}(k_2)x-k_2x},\\
&& \nonumber { E}^{(3)}_{2x}=\left[\frac{i\beta_0}{k_2}A^{(3)}_2-\frac{i\beta_0}{2k_2^3}L_{2z}-\frac{i\beta_0}{2k_2^2}L_{2z}x+\frac{L_{2x}}{k_2^2}\right]e^{-k_2x}\\
&& \,\,\,\,\,\,\,\,\,\,\,\,\,\,\,\,\,\,\,+\frac{1}{k^2_2}\left[N_{x}+\frac{i\beta_0N_{z}\left[2{\rm Re}(k_2)+k_2\right]}{4{\rm Re}(k_2)\left[{\rm Re}(k_2)+k_2\right]}\right]e^{-2{\rm Re}(k_2)x-k_2x},
\end{eqnarray}
where $A^{(3)}_{\alpha}$ ($\alpha=1,2$) are envelope functions (their concrete forms are not needed), and
\begin{eqnarray}
&& N_{x}=k_0^2\mu_2\frac{\beta_0^2+|k_2|^2}{|k_2|^2}\chi^{(3)}_{a0}|A|^2A,\\
&& N_{z}=-\frac{\left(\beta_0^2+|k_2|^2\right)\left[2\beta_0^2{\rm Re}(k_2)+\varepsilon_{20}\mu_2k_0^2k_2\right]}{\varepsilon_{20}k_2|k_2|^2}\chi^{(3)}_{a0}|A|^2A.
\end{eqnarray}
The expressions of $E_{\alpha y}^{(3)}$  ($\alpha=1,2$) are not needed and thus omitted here.


\vspace{5mm}
\noindent\textbf{\Large \sf Acknowledgments}\\
\noindent This work was supported by the NSF-China under Grants No.~11475063 and No.~11474099¡£

\vspace{5mm}
\noindent\textbf{\Large \sf Author contributions}\\
\noindent Q.Z. and C.T. carried out the calculations and wrote the manuscript. G.H. conceived the idea, conducted the calculations and revised the manuscript.

\vspace{5mm}
\noindent\textbf{\Large \sf Additional information}\\
Competing financial interests: The authors declare no competing financial interests.

\end{document}